\documentclass[showpacs,amsmath,amssymb,nofootinbib,prd]{revtex4}
\usepackage{graphicx}
\usepackage{enumerate}
\usepackage{amsmath}
\usepackage{amssymb}
\usepackage{amsfonts}
\usepackage{calligra}
\usepackage{color}
\usepackage{url}
\usepackage{hyperref}
\usepackage{subfigure}
\usepackage{mathtools}

\newcommand{\be}{\begin{equation}}
\newcommand{\ee}{\end{equation}}
\newcommand{\bn}{\begin{eqnarray}}
\newcommand{\en}{\end{eqnarray}}
\newcommand{\eps}{\epsilon}

\begin{document}

\title{Singlet scalar Dark Matter in Dark Two Higgs Doublet Model }

\author{ R. Gait\'an}
\email{rgaitan@unam.mx} \affiliation{Centro de Investigaciones Te\'oricas, FES Cuatitl\'an, UNAM, Apartado Postal 142, Cuatitl\'an-Izcalli, Estado de M\'exico, C\'odigo postal 54700, M\'exico.}
\author{E. A. Garc\'es}
\email{egarces@fis.cinvestav.mx}\affiliation{Centro de Investigaciones Te\'oricas, FES Cuatitl\'an, UNAM, Apartado Postal 142, Cuatitl\'an-Izcalli, Estado de M\'exico, C\'odigo postal 54700, M\'exico.}
\author{J. H. Montes de Oca}
\email{jmontes@fis.cinvestav.mx} \affiliation{Departamento de F\'{\i}sica, Centro de Investigaci\'on y Estudios Avanzados del I.P.N., Apdo. Postal 14-740, M\'{e}xico D.F., 07000, M\'{e}xico }

\begin{abstract}
We consider the case of the Dark Two Higgs Doublet Model (D2HDM) where a $U(1)'$ symmetry group and an extra Higgs doublet are added to the Standard Model.
This model leads to a gauge singlet particle as an interesting Dark Matter (DM) candidate.
We obtain phenomenological constraints to the parameter space of the model considering the one necessary to
produce the correct density of thermal relic dark matter  $\Omega h^2$.
We find a relation between the masses of the DM matter candidate $m_S$ and $m_{Z'}$ that satisfy the relic density for
given values of $\tan\beta$.
\end{abstract}


\maketitle

\section{Introduction}\label{sec:int}

The existence of Dark matter (DM) is now essentially established~\cite{Bertone:2004pz}.
The most convincing evidence for DM came from the observation that luminous objects
such as stars, gas clouds, globular clusters, or entire galaxies move faster than one
would expect if they only felt the gravitational attraction of other visible objects~\cite{Roos:2010wb,dark_theory}.
DM is estimated to constitute about 23~$\%$ of the total matter in the universe.
However, the origin of DM still remains a mystery.

The Standard Model (SM) of particle physics explains experimental results,  but none
of the SM particles can be a good candidate for the dark matter. Therefore, it is necessary
to look for new physics beyond the SM. In the literature there are various proposals as dark matter
candidates, being a weakly interacting massive particle (WIMP) a promising candidate; in fact,
the WIMP relic density is around the observed value (Planck Collaboration)~\cite{Ade:2013zuv}:

\begin{equation}
\Omega h_{CDM}^2 = 0.1199 \pm 0.0027.
\end{equation}\label{eq:Omega}
In supersymmetric models, one candidate is the lightest neutralino as a spin-1/2
WIMP dark matter~\cite{Jungman:1995df}. Spin-1 WIMP dark matter has been studied in the context of models with
extra dimensions~\cite{Servant:2002aq}. Some extensions of the SM include one or more
gauge singlet scalars~\cite{McDonald:1993ex,Silveira:1985rk,Burgess:2000yq}; if these scalars are stable, they can in principle account for a density
of DM.

Several new physics scenarios with extended scalar fields and extra Higgs doublets have been considered to accommodate
new particles as DM candidates.
For example, in the Inert Two Higgs Doublet Model (IDM)~\cite{Dolle:2009fn}, where a DM candidate coming from the Higgs doublet is considered, and its mass has been constrained
to the order of GeV~\cite{LopezHonorez:2006gr,LopezHonorez:2010tb}. 
Recently a two Higgs doublet model (2HDM) with a $U(1)$ gauge symmetry, including a
gauge boson $Z^{\prime}$ with mass of order of GeV scale or below, has been proposed, that is the case of the
D2HDM~\cite{Davoudiasl:2012ag,Lee:2013fda}.

In the D2HDM a continuous symmetry $U(1)$ is the mechanism to keep very suppressed the Yukawa couplings between fermions and scalars, while in the IDM this is achieved by imposing a $Z_2$ discrete symmetry. In this work we consider the framework of the D2HDM
in order to propose a WIMP-like DM candidate particle.

In section \ref{sec:mod} we will introduce the D2HDM, in section \ref{sec:DM} we discuss the relic density contribution of  a singlet scalar DM particle, and finally in section \ref{sec:res} we discuss the phenomenological limits obtained from cosmological data to the parameter space of the model.

\section{The model}\label{sec:mod}

In this section we will briefly introduce the field content and the parameters used later.
We will follow closely the notation introduced by Lee and Sher~\cite{Lee:2013fda}.

The gauge group of the D2HDM is the SM group extended with the $U(1)'$ group, $G_{D2HDM}=G_{SM}\times U(1)'$.
The coupling associated to $U(1)'$ will be denoted as $g_{Z'}$.  A mixing among the kinetic energy terms for
the gauge fields, $U(1)_{Y}$ and $U(1)'$, is allowed given the gauge invariance in $G_{D2HDM}$. This mixing
term is parametrized by the weak mixing angle $\theta_W$ and a new parameter denoted by $\varepsilon$.
The kinetic terms are written explicitly
as
\begin{eqnarray}
\mathcal{L}_{Kin}& =& - \frac{1}{4} \hat{B}_{\mu\nu} \hat{B}^{\mu\nu}+ \frac{1}{2}\frac{\varepsilon}{\cos\theta_{W}} \hat{B}^{\mu\nu} \hat{Z'}_{0\mu\nu}\\ \nonumber
&-& \frac{1}{4}\hat{Z'}_{0\mu\nu}\hat{Z'_0}^{\mu\nu},\label{eq:kin}
\end{eqnarray}
where, $ \hat{B}^{\mu\nu}$ and  $\hat{Z'_0}^{\mu\nu}$ are the field strength tensors of the  $U(1)_Y$ and $U(1)'$ gauge bosons, respectively.
The gauge field of $Z'$ is redefined by the rotation
\begin{eqnarray}
 \left( \begin{array}{c}
 Z'_{0\mu}\\
B_\mu
\end{array} \right)
=
 \left( \begin{array}{cc}
\sqrt{1-\varepsilon^2/\cos^2\theta_W} & 0\\
-\varepsilon/ \cos^2\theta_W&1 \end{array} \right) \left( \begin{array}{c}
 \hat{Z'}_{0\mu}\\
\hat{B}_\mu
\end{array} \right),
\end{eqnarray}
and therefore the mixing term in Equation (\ref{eq:kin}) is cancelled. The mixing parameter $\varepsilon$ will  appear
 in all terms where $Z'$ is redefined. The magnitude of $\varepsilon$ has been constrained to $\varepsilon\leq10^{-3}$ \cite{Davoudiasl:2012ag,eps1,holdom}.

In the D2HDM the scalar field content is given in the two doublets and one singlet defined as follows

\begin{eqnarray}
\Phi_1 &=& \left( \begin{array}{c}
\phi_1^+\\
\frac{1}{\sqrt{2}}(v_1+\phi_1+i\eta_1)
\end{array} \right),\\ \nonumber
\Phi_2 &=& \left( \begin{array}{c}
\phi_2^+\\
\frac{1}{\sqrt{2}}(v_2+\phi_2+i\eta_2)
\end{array} \right),\\ \nonumber
\Phi_s &=& \frac{1}{\sqrt{2}}(v_s+\phi_s+i\eta_s),
\label{eq:scalar_reps}
\end{eqnarray}
where $v_1, v_2$ and $v_s$ are the vacuum expectation values (VEV), $v_1$ and  $v_2$ satisfy $v^2=v_1^2+v_2^2$, with $v=246$~GeV.  The charges and representations for the scalar fields under the symmetry group $G_{D2HDM}$
can be written in compact form as follows
\begin{eqnarray}
\Phi_1\sim(1, 2, 1/2, 0),\\ \nonumber
\Phi_2\sim(1,2,1/2,1),\\ \nonumber
\Phi_s\sim(1,1,0,1),
\end{eqnarray}
where the quantum numbers are in the following order $(SU(3), SU(2),Y, Q')$.

The interactions between the scalar and gauge bosons are given by
\begin{eqnarray}
\mathcal{L}_\text{scalar} &=& | D_\mu \Phi_1 |^2 + | D_\mu \Phi_2 |^2 + | D_\mu \Phi_S |^2,
\end{eqnarray}
where the covariant derivative $D_\mu$ is defined as
\begin{eqnarray}
D_\mu =  \left(  \partial_\mu + i g' Y \hat B_\mu + i g T_3 \hat W_{3 \mu} + i g_{Z'} Q' \hat Z'_{0\mu} \right) .
\end{eqnarray}
After the right spontaneous symmetry breaking SSB, additionally to the mass terms we will have the mixing term
$(Z_{0\mu} Z_0'^{\mu})$.
This new mixing term is proportional
to $\Delta^2=\frac{1}{2} g_{Z'} g_{Z} v^2 \cos^2\beta + \frac{1}{4} \frac{\varepsilon}{\cos\theta_W} g_{Z} g' v^2$,  and the $Z'$ mass is
\begin{eqnarray}
m_{Z'^0}^2 &=& g_{Z'}^2 (v^2 \cos^2\beta + v_S^2) + \frac{\eps}{\cos\theta_W} g_{Z'} g' v^2 \cos^2\beta + \frac{1}{4} \left(\frac{\eps}{\cos\theta_W}\right)^2 g'^2 v^2 . \label{eq:mZ'}
\end{eqnarray}
In order to cancel the mixing term the following rotation is required
\begin{eqnarray}
\left(  \begin{array}{c}
Z \\ Z'
\end{array} \right)
=
\left( \begin{array}{cc}
\cos\xi & -\sin\xi \\ \sin\xi & \cos\xi
\end{array} \right)
\left(
\begin{array}{c}  Z^0 \\ Z'^0
\end{array} \right),
\end{eqnarray}
where the mixing angle $\xi$ satisfy the expression $\tan 2\xi = \frac{2\Delta^2}{m^2_{Z^0}-m^2_{Z^{'0}}}$, and has been constrained to $|\xi|<10^{-3}$~\cite{Bouchiat:2004sp}.

The most general potential for the scalar sector is given by
\begin{eqnarray}
V=V_{1}+V_{2}+V_{3},
\end{eqnarray}
where
\begin{eqnarray}
V_{1} &=&\mu _{1}^{2}\Phi _{1}^{\dag }\Phi _{1}+\mu _{2}^{2}\Phi _{2}^{\dag
}\Phi _{2}+\frac{1}{2}\lambda _{1}\left( \Phi _{1}^{\dag }\Phi _{1}\right)
^{2}+\frac{1}{2}\lambda _{2}\left( \Phi _{2}^{\dag }\Phi _{2}\right) ^{2} \\
&&+\lambda _{3}\left( \Phi _{1}^{\dag }\Phi _{1}\right) \left( \Phi
_{2}^{\dag }\Phi _{2}\right) +\lambda _{4}\left( \Phi _{1}^{\dag }\Phi
_{2}\right) \left( \Phi _{2}^{\dag }\Phi _{1}\right) ,
\label{V1}
\end{eqnarray}
\begin{eqnarray}
V_{2}=\mu _{3}^{2}\Phi _{s}^{\dag }\Phi _{s}+\frac{1}{2}\lambda _{6}\left(
\Phi _{s}^{\dag }\Phi _{s}\right) ^{2},
\label{V2}
\end{eqnarray}
\begin{eqnarray}
V_{3}=\lambda _{7}\left( \Phi _{1}^{\dag }\Phi _{1}\right) \left( \Phi
_{s}^{\dag }\Phi _{s}\right) +\lambda _{8}\left( \Phi _{2}^{\dag }\Phi
_{2}\right) \left( \Phi _{s}^{\dag }\Phi _{s}\right).
\label{V3}
\end{eqnarray}
We note that all parameters in the potential are real parameters. Complex parameter could arise from terms like
$\Phi
_{1}^{\dag }\Phi _{2}$ or $(\Phi
_{1}\Phi _{2})^2$, we do not have this kind of terms because they are forbidden by the $U(1)^\prime$ gauge symmetry.

The mass squared matrices for the Higgs and singlet fields are
\begin{equation}
M_{Higgs}^{2}=M_{H^{\pm }}^{2}\oplus m_{H,h}^{2}\oplus M_{s}^{2},
\end{equation}
where
\begin{equation}
M_{H^{\pm }}^{2}=\frac{1}{2}\lambda _{4}\left(
\begin{array}{cc}
-v_{2}^{2} & v_{1}v_{2} \\
v_{1}v_{2} & -v_{1}^{2}%
\end{array}%
\right) ,
\end{equation}
\begin{equation}
M_{H,h}^{2}=\left(
\begin{array}{cc}
\lambda _{1}v_{1}^{2} & \left( \lambda _{3}+\lambda _{4}\right) v_{1}v_{2}
\\
\left( \lambda _{3}+\lambda _{4}\right) v_{1}v_{2} & \lambda _{2}v_{2}^{2}%
\end{array}%
\right) ,
\end{equation}
\begin{equation}
M_{s}^{2}=\left(
\begin{array}{cc}
\lambda _{6}v_{s}^{2} & 0 \\
0 & 0%
\end{array}%
\right) .
\end{equation}
Once the mass matrices are diagonalized, the Higgs bosons masses are found to be
\begin{eqnarray}
m_{H}^{2}= \frac{1}{2} \left( \lambda_1 v_1^2 + \lambda_2 v_2^2 + \sqrt{ (\lambda_1 v_1^2 - \lambda_2 v_2^2)^2 + 4 (\lambda_3 + \lambda_4)^2 v_1^2 v_2^2} \right),
\end{eqnarray}
\begin{eqnarray}
m_{h}^{2}= \frac{1}{2} \left( \lambda_1 v_1^2 + \lambda_2 v_2^2 - \sqrt{ (\lambda_1 v_1^2 - \lambda_2 v_2^2)^2 + 4 (\lambda_3 + \lambda_4)^2 v_1^2 v_2^2} \right),
\end{eqnarray}
\begin{eqnarray}
m_{H^\pm}^2 = -\frac{\lambda_4}{2} v^2,
\end{eqnarray}
\begin{eqnarray}
m_{s}^{2}= \lambda_6 v_S^2.\label{eq:ms}
\end{eqnarray}
The mixing angle $\alpha$ is constrained by
\begin{eqnarray}
\tan \left( 2\alpha \right) =\frac{2\left( \lambda _{3}+\lambda _{4}\right)
v_{1}v_{2}}{\lambda _{1}v_{1}^{2}-\lambda _{2}v_{2}^{2}}.
\end{eqnarray}
The potential $V_3$ will be neglected in order to avoid interactions between the scalar doublets and the singlet.
Note that $\lambda_7$ and $\lambda_8$ do not appear in the mass matrix, even if $V_3$ is included, this is due to the minimum condition for the potential.

In the usual 2HDM, the pseudoscalar mass $m_{A^0}$ is proportional to the parameter $\lambda_5$ associated to the $(\Phi_1^{\dagger}\Phi_2)^2$ term. In the D2HDM, $\lambda_5=0$ in order to keep the $U(1)^\prime$ gauge invariance, which means that initially the pseudoscalar does not acquire mass through the
usual potential.
The $\lambda_5$ coupling can be generated by introducing the interaction of the form $(\Phi_1^\dagger\Phi_2)^2\Phi_s^2$ between the doublets and the singlet~\cite{Ko:2013zsa}, however this issue is beyond the scope of our work and will not be taken into account.

\section{Relic abundance of the Singlet as Dark Matter candidate}\label{sec:DM}
In  D2HDM the two doublet scalars $\Phi_{1,2}$  have interactions with fermions, therefore none of them could be proposed as dark matter candidates.
However the singlet scalar $\Phi_s$ has no Yukawa interactions, playing the role of inert particle in the D2HDM.

We assume that DM particles were initially in thermal equilibrium with the early universe. After the temperature of the universe
goes down, below the DM particle mass scale, the expansion rate overtakes the annihilation rate for the DM particle. In other words the
density number is no longer affected by interactions and remains constant. This is the so called freeze out mechanism.
The density of a specific particle at the time of freeze out is known as the relic density for this particle~\cite{Dodelson:2003ft}.

\begin{centering}
\begin{figure}
\subfigure[]{\includegraphics[width=4.cm]{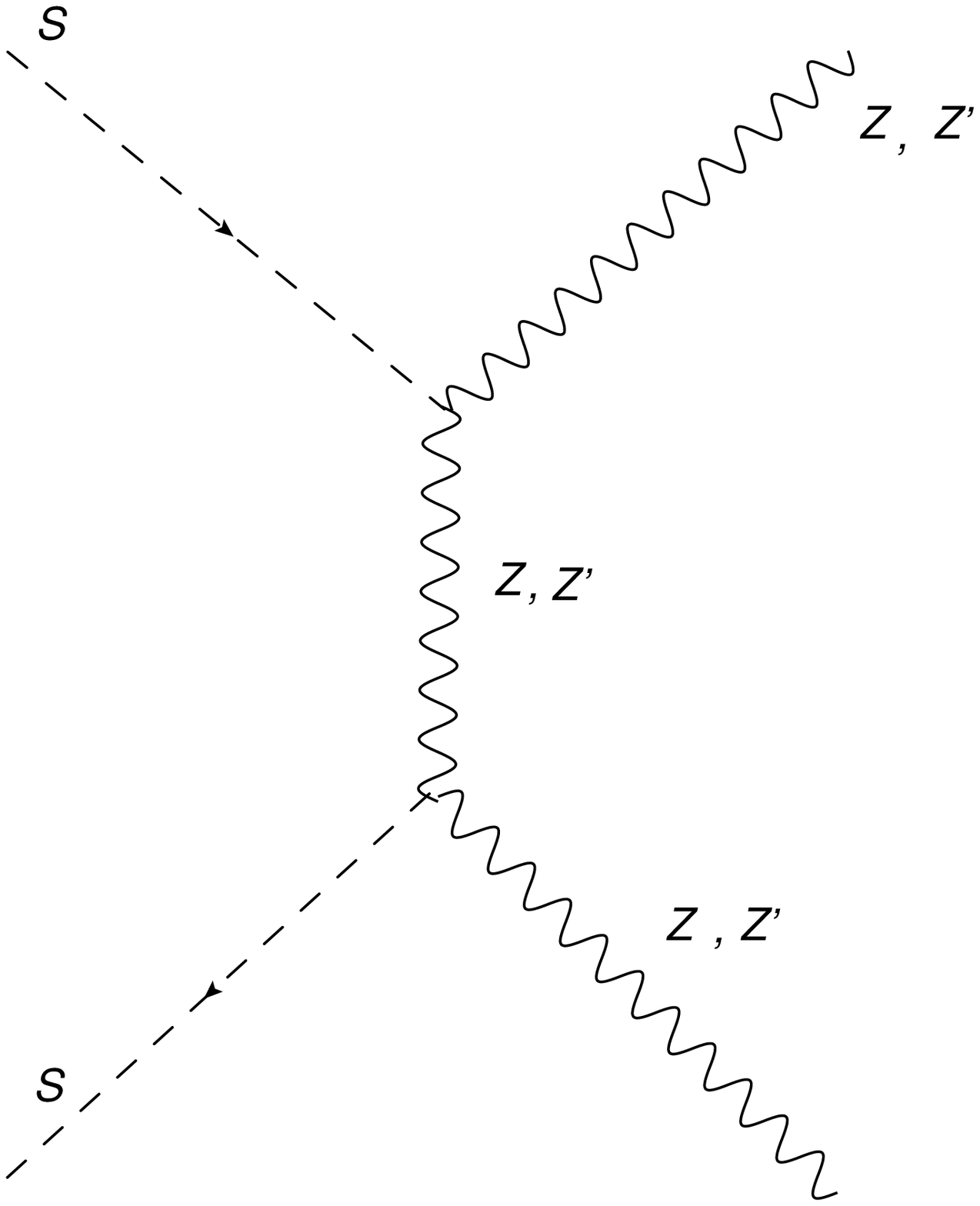}}
\subfigure[]{\includegraphics[width=6.5cm]{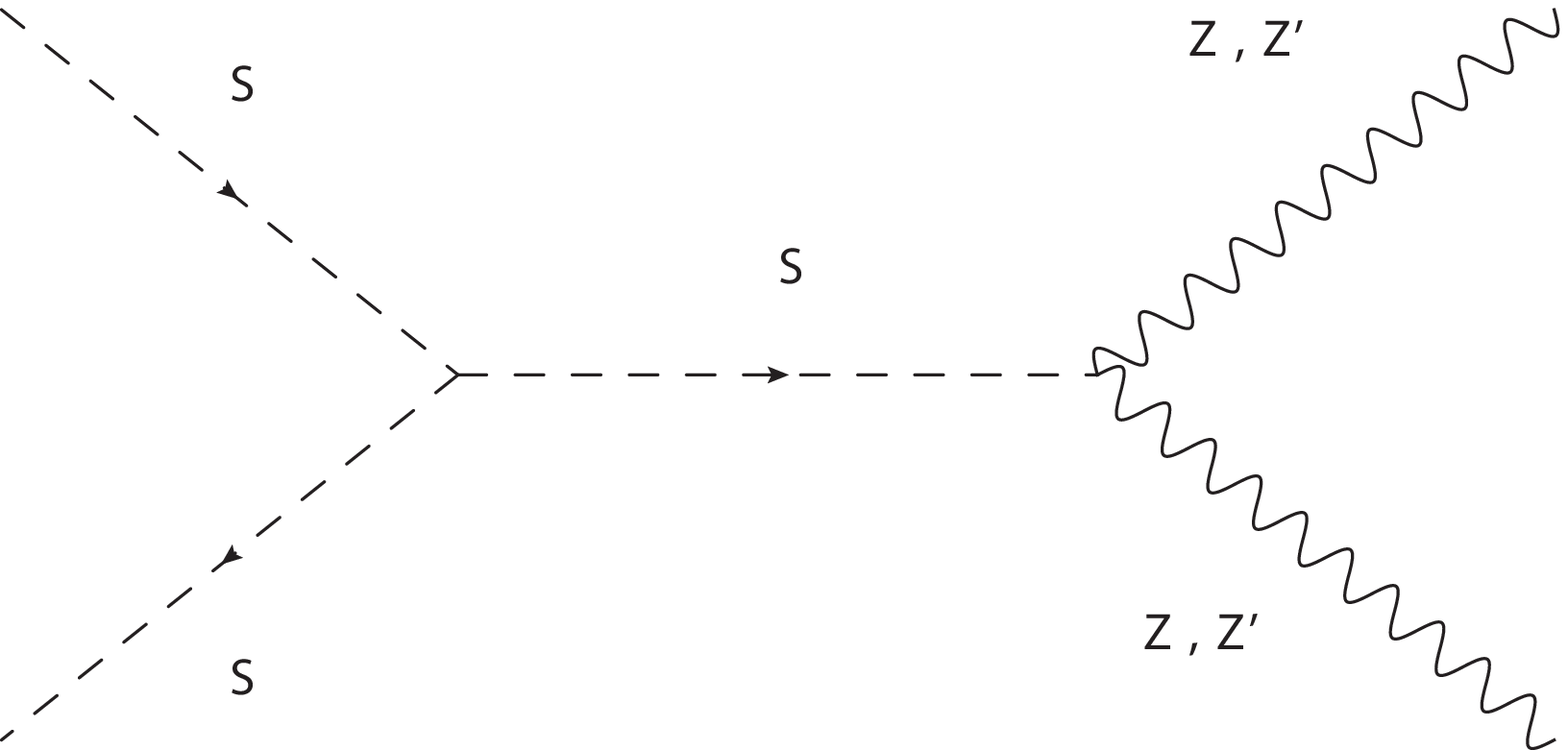}}
\subfigure[]{\includegraphics[width=4.5cm]{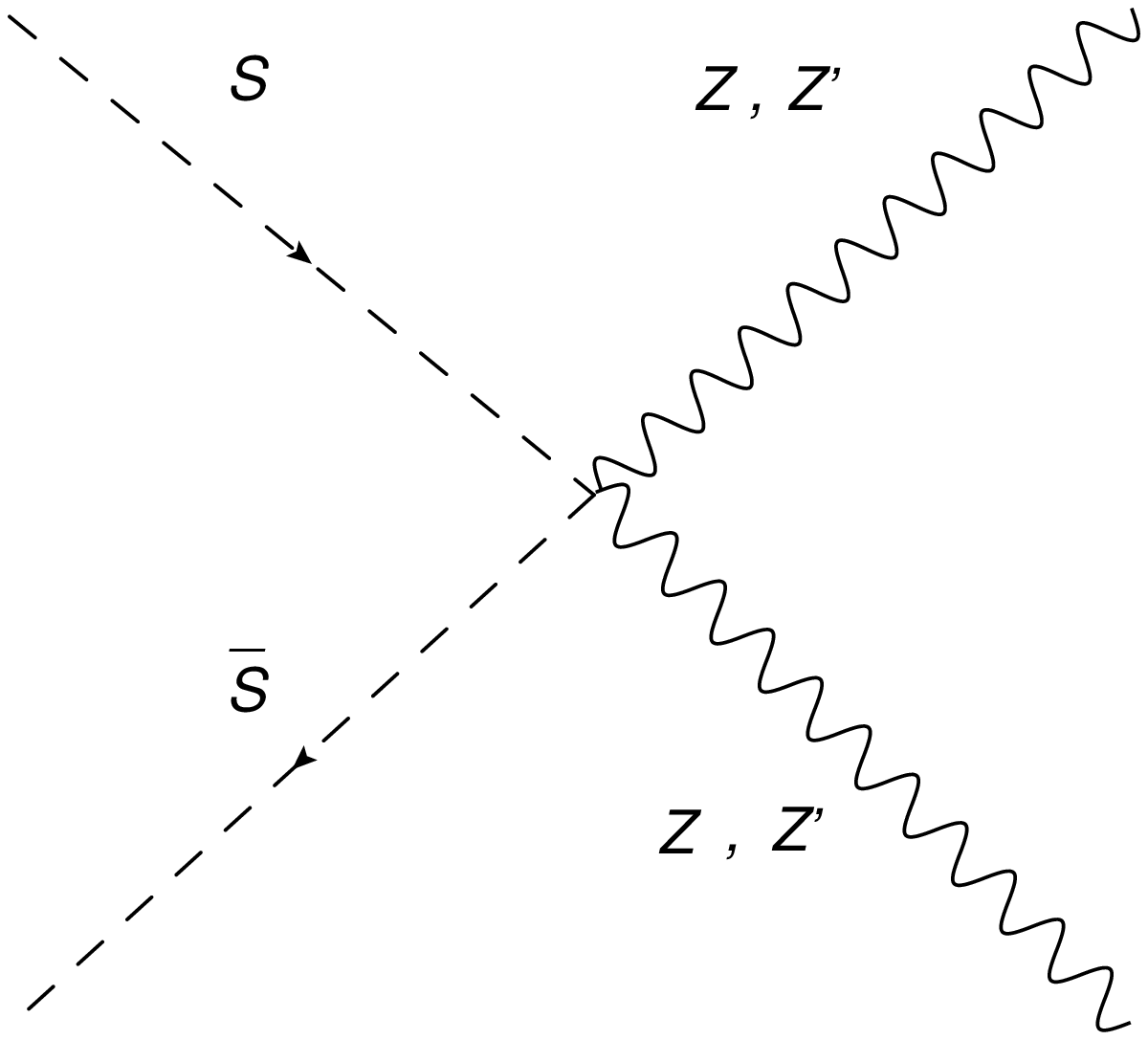}}
\caption{Feynmann diagrams of the annihilation channels with $Z$ and $Z'$ bosons in the final states. }\label{fig:02}
\end{figure}
\end{centering}

The Dark matter abundance can be obtained by solving the Boltzmann equation for the number density rate. The Boltzmann equation can be written as
\begin{equation}
a^{-3}\frac{d}{dt}\left( na^{3}\right) =\left\langle \sigma v\right\rangle
\left( n_{eq}^{2}-n^{2}\right),
\end{equation}
where $n$ is the DM number density  and $a$ is a scale factor. The Species-Independent equilibrium number density is defined as
\begin{equation}
n_{eq}=\left\{
\begin{array}{cc}
g_{s}\left( \frac{m_{s}T}{2\pi }\right) ^{\frac{3}{2}}e^{-\frac{m_{s}}{T}},
& m_{s}>>T \\
g_{s}\frac{T^{3}}{\pi ^{2}}, & m_{s}<<T%
\end{array}%
\right. ,
\end{equation}
where $g_s$ is the degeneracy of the scalar DM.  All information about the model interactions is contained in the thermally averaged cross section, defined as
\begin{equation}
\left\langle \sigma v\right\rangle =\frac{\left( 2\pi \right) ^{4}}{%
(n_{eq})^2}\prod\limits_{i=1}^{4}\int \frac{d^{3}p_{i}}{2E_{i}}%
e^{-\left( E_{1}+E_{2}\right) /T}\delta ^{4}\left(
p_{1}+p_{2}-p_{3}-p_{4}\right) \left\vert \mathcal{ M}\right\vert ^{2}.
\end{equation}
In Fig. (\ref{fig:02}) are shown the annihilation channels contributing to the total amplitude in the model.
The total average amplitude $|\mathcal{M}|^2$ is
\begin{equation}
|\mathcal{M}|^2= |\mathcal{M}_{ZZ}|^2+|\mathcal{M}_{Z'Z'}|^2+|\mathcal{M}_{ZZ'}|^2,
\end{equation}
where
\begin{equation}
\left\vert \mathcal{M}_{ZZ}\right\vert ^{2}=4g_{z^{\prime }}^{4}\sin ^{4}\xi \left(
\frac{2g_{z^{\prime }}^{2}v_{S}^{2}\sin ^{2}\xi }{s-m_{Z}^{2}}+\frac{%
2g_{z^{\prime }}^{2}v_{S}^{2}\cos ^{2}\xi }{s-m_{Z^{\prime }}^{2}}-\frac{%
6m_{S}^{2}}{s-m_{S}^{2}}+1\right) ^{2}\left[ 2+\frac{1}{4m_{Z}^{4}}\left(
s-2m_{Z}^{2}\right) ^{2}\right] ,
\label{eq:zz}
\end{equation}
\begin{equation}
\left\vert \mathcal{M}_{Z^{\prime }Z^{\prime }}\right\vert ^{2}=4g_{z^{\prime
}}^{4}\cos ^{4}\xi \left( \frac{2g_{z^{\prime }}^{2}v_{S}^{2}\cos ^{2}\xi }{%
s-m_{Z^{\prime }}^{2}}+\frac{2g_{z^{\prime }}^{2}v_{S}^{2}\sin ^{2}\xi }{%
s-m_{Z}^{2}}-\frac{6m_{S}^{2}}{s-m_{S}^{2}}+1\right) ^{2}\left[ 2+\frac{1}{%
4m_{Z^{\prime }}^{4}}\left( s-2m_{Z^{\prime }}^{2}\right) ^{2}\right] ,
\label{eq:zz'}
\end{equation}
\begin{equation}
\left\vert \mathcal{M}_{ZZ^{\prime }}\right\vert ^{2}=4g_{z^{\prime }}^{4}\cos ^{2}\xi
\sin ^{2}\xi \left( \frac{2g_{z^{\prime }}^{2}v_{S}^{2}\sin ^{2}\xi }{%
s-m_{Z}^{2}}+\frac{2g_{z^{\prime }}^{2}v_{S}^{2}\cos ^{2}\xi }{%
s-m_{Z^{\prime }}^{2}}-\frac{6m_{S}^{2}}{s-m_{S}^{2}}+1\right) ^{2}\left[ 2+%
\frac{1}{4m_{Z}^{2}m_{Z^{\prime }}^{2}}\left( s-m_{Z}^{2}-m_{Z^{\prime
}}^{2}\right) ^{2}\right] ,
\label{eq:z'z'}
\end{equation}
where $s$ is the Mandelstam variable for the squared center of mass energy.
From the equations above, it can be noticed that the amplitud $\left\vert \mathcal{M}_{Z^{\prime}Z^{\prime}}\right\vert ^{2}$ gives the biggest contribution to the total amplitude due to the
strong dependence in the mixing parameter $\cos ^{4}\xi$.

Interactions derived from $\Phi_s$ are under control by the condition $m_s<2m_X$, being $X$ any gauge neutral boson in the model.
Under these assumptions the scalar particle associated to the $\Phi_s$ singlet  is proposed as a DM candidate.

\section{Results and discussion}\label{sec:res}

In previous sections was stablished the motivation to propose $\Phi_s$ as a DM candidate. In the following we will
show how the model parameters can be strongly constrained by taking into account the latest measurements
of relic density.

In order to compute the relic density for our Dark Matter particle, given different values of the
parameters in the model, the Boltzmann equation was solved numerically using micrOMEGAs~\cite{Belanger:2008sj,Belanger:2013oya}.
The squared matrix elements for the New Physics model file were generated through LanHEP~\cite{Semenov:2008jy}, implementing the code in order to include the
D2HDM.

\begin{figure}
\hfill
\subfigure[]{\includegraphics[width=8.9cm]{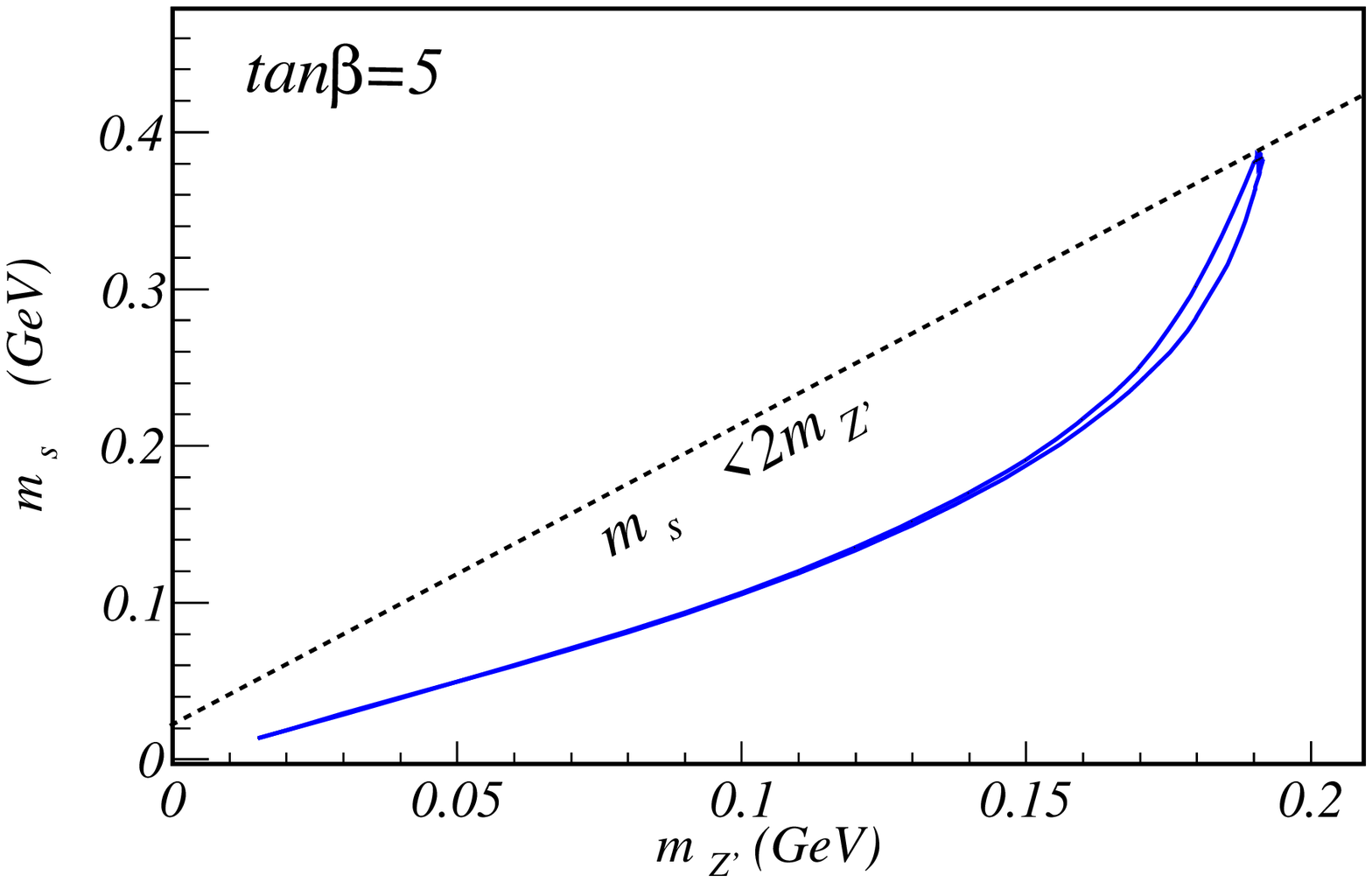}}
\hfill
\subfigure[]{\includegraphics[width=8.9cm]{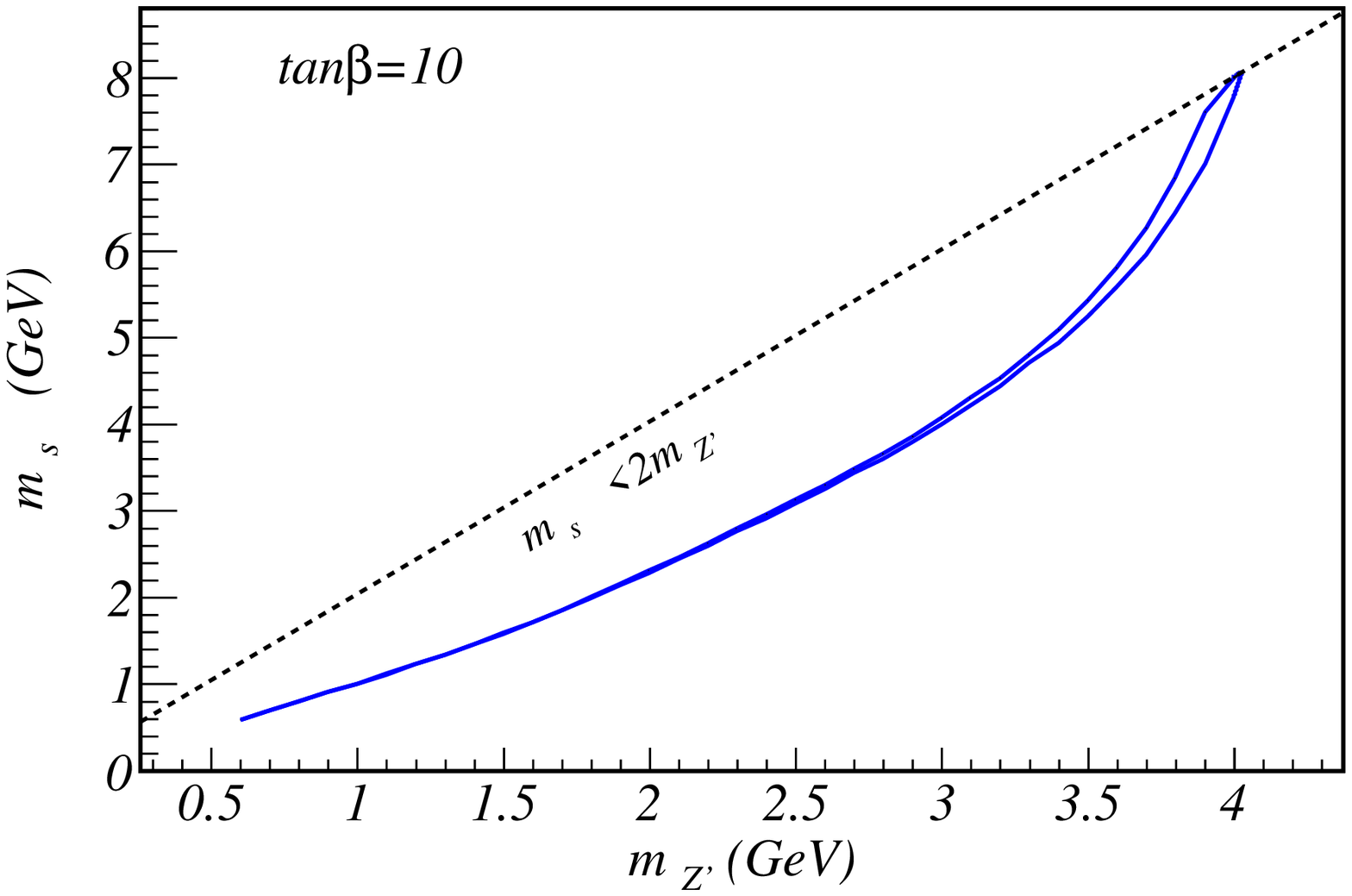}}
\hfill
\caption{$m_s$ as a function of $m_{Z'}$ necessary to produce the DM relic density, $\Omega h^2=0.1199\pm0.0027$, when $m_h=60$~GeV and (a) $\tan\beta=5$, (b)$\tan\beta=10.$}\label{fig:00}
\end{figure}
\begin{figure}
\hfill
\subfigure[]{\includegraphics[width=8.9cm]{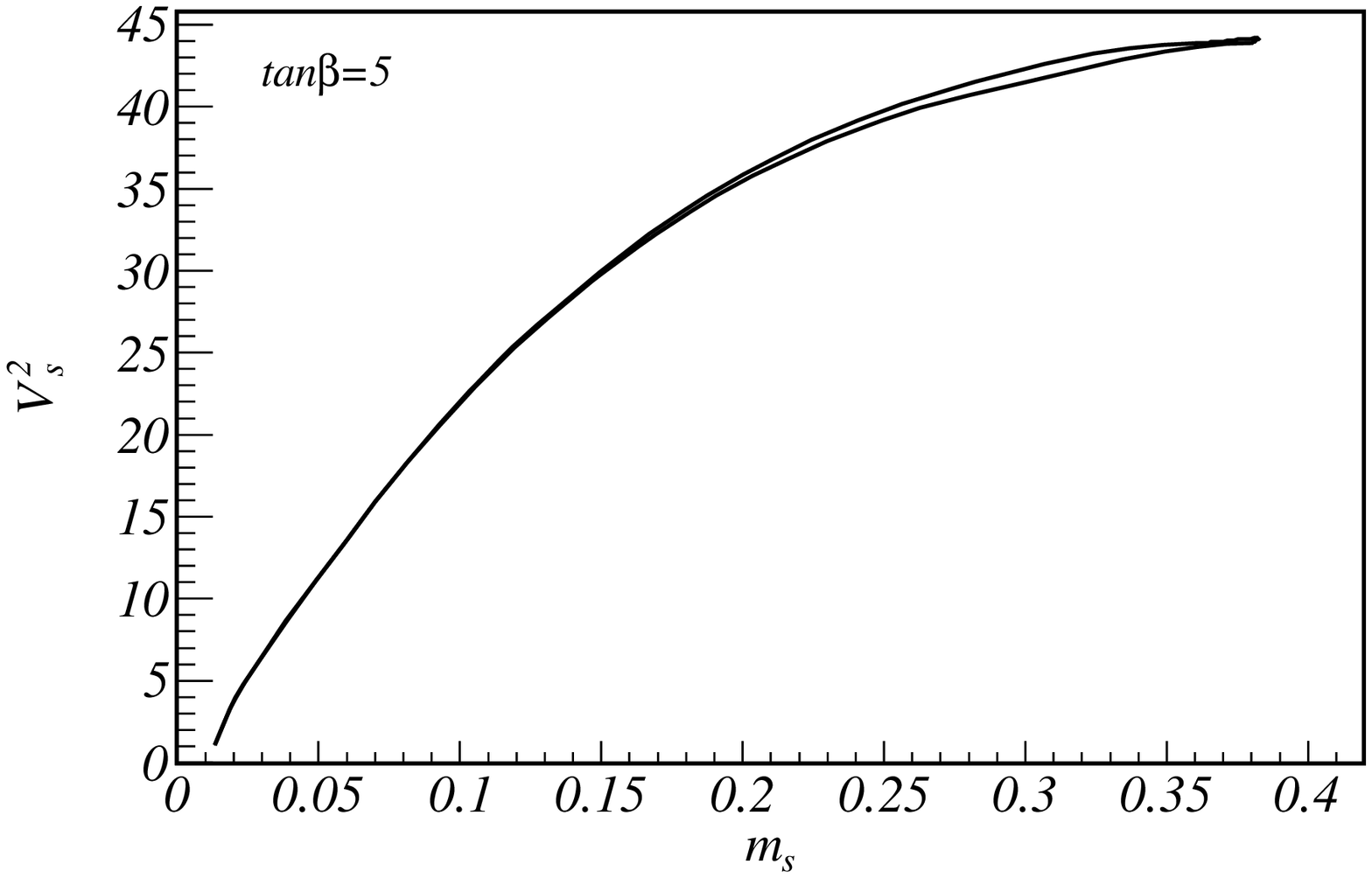}}
\hfill
\subfigure[]{\includegraphics[width=8.9cm]{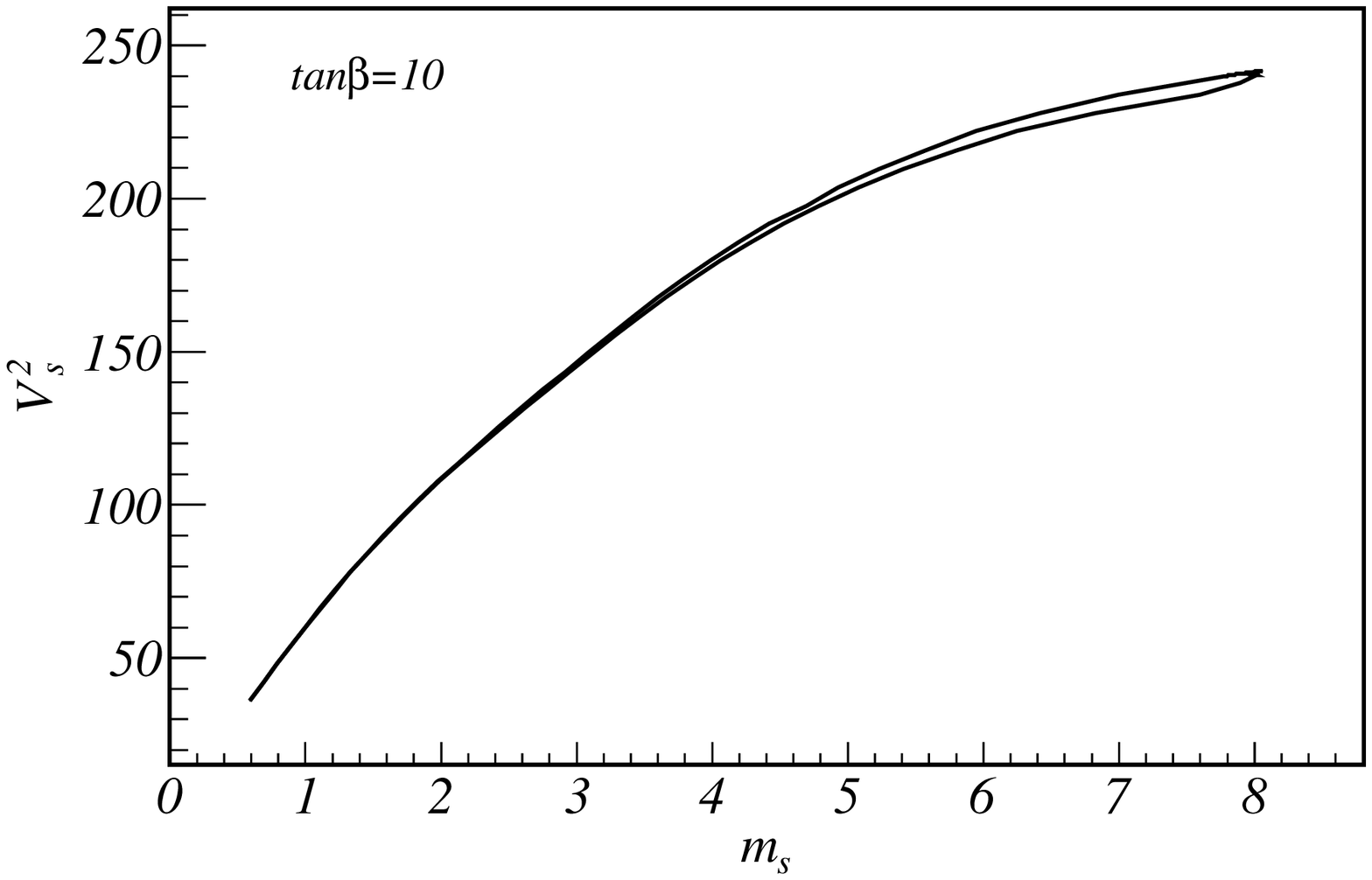}}
\hfill
\caption{$V_s^2$~(GeV$^2$) as a function of $m_s$~(GeV), corresponding to the a velue of DM relic density of, $\Omega h^2=0.1199\pm0.0027$, for $m_h=60$~GeV and (a) $\tan\beta=5$, (b)$\tan\beta=10$.}\label{fig:01}
\end{figure}

The charged and neutral Higgs masses are used to evaluate the $\lambda$ parameters that appear in the potential, equation (\ref{V1}). The case under study is that of a neutral Higgs lighter than the one observed by ATLAS and CMS collaborations~\cite{Aad:2012tfa,Chatrchyan:2012ufa}.
For our calculations the heavier neutral Higgs mass is fixed to $m_{H}=125$ GeV and the mass of the smaller Higgs varies
in a range  below that value.

The case of a neutral Higgs, with mass lighter than $125$~GeV, implies that given the smallness of its branching ratio it cannot be observed in collision experiments, this is the case of the so called
invisible Higgs~\cite{Djouadi:2012zc}.
The mixing angles of the model $\alpha$ and $\beta$ are fixed by the mass region for the invisible Higgs.
Both are phenomenologically constrained by data of the invisible Higgs dacay to fermions from LEP~\cite{Achard:2004cf,Abdallah:2003ry,Abbiendi:2007ac}.

In Fig. (\ref{fig:00}) is depicted the allowed region for the mass of the scalar DM particle, $m_s$, and the mass of the light $Z^{\prime}$, $m_{Z'}$.
In Fig. (\ref{fig:01}) it is shown $v_S^2$   as a function of $m_S$. In both cases the narrow region is bounded by the value of the relic density
in the range of the measured value $\Omega h^2=0.1199\pm0.0027$~\cite{Ade:2013zuv}. The mass of the invisible Higgs $m_h$ in all figures is set to $m_h=60$~GeV, and for different values in the range $[50 \leq m_h \leq 110]$~GeV, there is no significant difference in the results.
The allowed values obtained depend  on a specific assignment of $\tan\beta$.

From Fig. (\ref{fig:00}) and Fig. (\ref{fig:01}) can be extracted the following limits,
for  $\tan\beta=5$, the restrictions found are $[0.02 \leq m_s \leq 0.4]$~GeV and
$[0.015\leq m_{Z'} \leq 0.2]$~GeV. And for $\tan\beta=10$,
we have $[0.5 \leq m_s \leq 8]$~GeV, and
$[0.6\leq m_{Z'}\leq4]$~GeV. Also for $v_s$ we have, $[1.0 \leq v_s \leq 6.6]$~GeV
when  $\tan\beta=5$, and $[6.7 \leq v_s \leq 15.5]$~GeV when $\tan\beta=10$.

The D2HDM includes a scalar spectrum originated from the two doublets and singlet, $H,h, H^{\pm},s$. The doublet participation in the Yukawa couplings is under control via the
$U(1)'$ gauge invariance, while the mixing angles $\alpha$ and $\beta$ are under control by the assumption of the invisible Higgs.
The singlet scalar can be a plausible WIMP-like DM candidate due to its lack
of participation in the Yukawa couplings.

From the amplitudes obtained for the processes $ss\rightarrow Z Z, Z Z' , Z'Z'$ in Eqs.
(\ref{eq:zz})-(\ref{eq:z'z'}), one can notice that the mixing parameters of the model also contribute in the constraining of the allowed values
of the vector boson and scalar particle masses $m_{Z'}$ and $m_s$.

\section{Conclusion}\label{sec:conc}

We find that the D2HDM provides a
plausible DM particle candidate. The masses of the scalar particle $m_s$
and the extra boson $m_{Z'}$, can be restricted using the
latest measurements of relic density from Microwave Background Radiation.
The model under study is a feasible one with a testable parameter space that could be further restricted using future data.


\section{Acknowledgments}
The authors thank Omar Miranda for useful discussions and comments to the manuscript.
This work has been supported by Conacyt grant 166639, by PAPIIT project IN117611-3, and by Sistema Nacional
de Investigadores (SNI), M\'exico.
E.A.G. thanks \textit{Programa de Becas Posdoctorales en la UNAM} and J.H.M.O. is thankful for the support from the postdoctoral CONACYT grant.

\end{document}